\documentclass[12pt]{article}
\usepackage[utf8]{inputenc}

\usepackage[left=1in,right=1in,top=1in,bottom=1in]{geometry}
\usepackage{booktabs} 
\usepackage{authblk} 
\usepackage{graphicx} 
\usepackage{comment}
\usepackage{breakcites}
\usepackage[bookmarks=false]{hyperref}
\usepackage{amsmath}
\usepackage{amssymb}
\usepackage[dvipsnames]{xcolor}
\usepackage{lineno}

\title{Network efficiency of spatial systems with fractal morphology: a geometric graphs approach}

\author[1]{A. C. Flores-Ortega}
\author[2]{J. R. Nicolás-Carlock\footnote{Correspondence: jnicolas@unam.mx; carrillo@ifuap.buap.mx}}
\author[1]{J. L. Carrillo-Estrada$^*$}

\affil[1]{Instituto de Física, Benemérita Universidad Autónoma de México, México}
\affil[2]{Instituto de Física, Universidad Nacional Autónoma de México, México}
\date{}

\begin{document}

\maketitle


\begin{abstract}

The functional features of spatial networks depend upon a non-trivial relationship between the topological and physical structure. Here, we explore that relationship for spatial networks with radial symmetry and disordered fractal morphology. Under a geometric graphs approach, we quantify the effectiveness of the exchange of information in the system from center to perimeter and over the entire network structure. We mainly consider two paradigmatic models of disordered fractal formation, the Ballistic Aggregation and Diffusion-Limited Aggregation models, and complementary, the Viscek and Hexaflake fractals, and Kagome and Hexagonal lattices. First, we show that complex tree morphologies provide important advantages over regular configurations, such as an invariant structural cost for different fractal dimensions. Furthermore, although these systems are known to be scale-free in space, they have bounded degree distributions for different values of an euclidean connectivity parameter and, therefore, do not represent ordinary scale-free networks. Finally, compared to regular structures, fractal trees are fragile and overall inefficient as expected, however, we show that this efficiency can become similar to that of a robust hexagonal lattice, at a similar cost, by just considering a very short euclidean connectivity beyond first neighbors. 
\end{abstract}

\section*{Introduction}

The functional features of complex spatial systems depend upon a non-trivial relationship between the space-dependent structure (physical morphology) with the space-independent counterpart (network topology) \cite{barthelemy2022spatial}. Among the diverse physical structures found in natural and social systems, disordered fractals represent important systems of study due to the universality of their growth processes and morphological characteristics. For example, the fractal branching observed in living systems, such as neurons, bacterial colonies or hyphal networks, has been also observed in out-of-equilibrium physical phenomena, such as dielectric breakdown, viscous fingering, mineral deposition and colloidal aggregation \cite{meakin1998, vicsek1992, ben1990formation, Sander2011}. In particular, the growth dynamics of these physical phenomena fall into the Laplacian growth model, also known to belong to the \textsl{Diffusion Limited Aggregation} (DLA) universality \cite{Sander2011, nicolas2019}. The DLA fractal has well-defined fractal dimension according to embedding space \cite{witten1981, nicolas2019} and, due to the versatility of its pattern-formation mechanism, it has also been used for modelling complex spatial systems, from neurons \cite{luczak2010} to cities \cite{batty1989urban}. 

In this article we explore the non-trivial relationship between topology and morphology of spatial networks with radial symmetry and different degrees of disordered fractal morphology. Under a geometric graphs approach in which parts of the spatial system are connected via a euclidean connectivity parameter \cite{daqing2011, Papadopoulos2018, barthelemy2022spatial}, we quantify the effectiveness of exchange of information, from center to perimeter and over the entire network structure. Our analysis considers only two-dimensional systems of identical particles assembled by stochastic processes. The main models are the DLA and \textsl{Ballistic Aggregation} (BA) disordered fractals. For comparison purposes we also considered two deterministic structures, the \textsl{Vicsek} and \textsl{Hexaflake} fractals, and the non-fractal ordered systems, \textsl{Hexagonal} and \textsl{Kagome} lattices, constructed with similar radial characteristics. The analysis is divided into two parts: first, in a nearest neighbors euclidean approximation, we developed a structural characterization of the system's capabilities to explore the plane in terms of the \textsl{range} (the maximum linear extension with respect to the origin), the \textsl{coverage} (the ability to cover the plane in all directions), the \textsl{structural cost} (assembly connections), and \textsl{configurational complexity} (local connectivity); second, we quantified the effectiveness of exchange of information in terms of a \textsl{center-perimeter} communication ratio and the network \textsl{efficiency} (overall network communication) for the nearest neighbor approximation, then, we show how the efficiency and other properties of the system can be improved at a very low cost by leveraging on the geometric graphs approach beyond first-neighbors. Finally, we present a discussion on our main findings and their potential applications.

In the following, the term \textsl{morphology} refers to the physical or space-dependent structure (how things are distributed in space), while the terms \textsl{topology} or \textsl{network} refer to the space-independent structure (how things are connected with each other). In this way, the structure of spatial networks has both physical and topological properties \cite{stephenson2017topological, naoya2018}.

\section*{Results}

\begin{figure}[t!]
\includegraphics[width=1.0\textwidth]{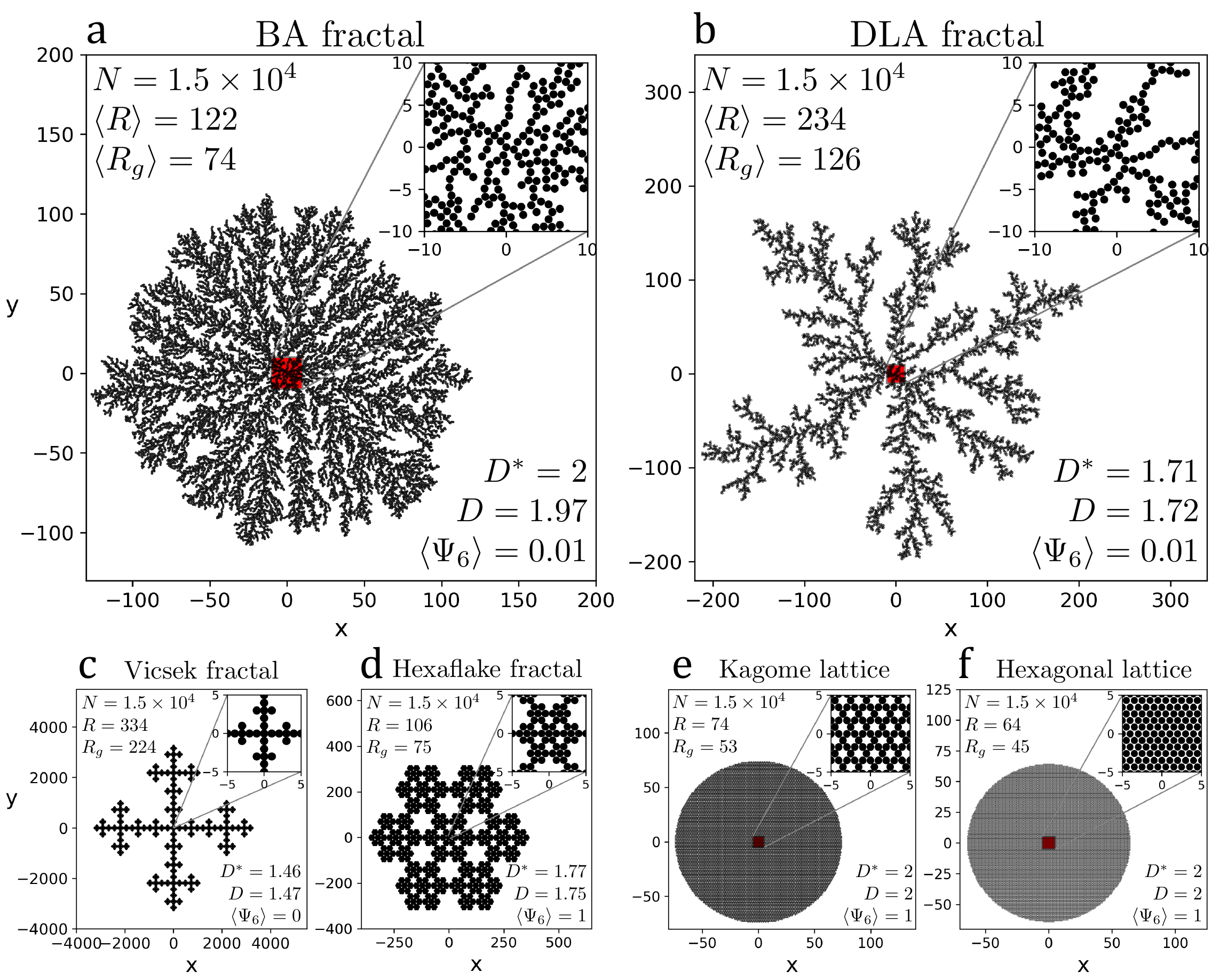}
\caption{\small \textbf{Models.} Stochastic fractals: (a) Ballistic Aggregation (BA) and (b) Diffusion-Limited Aggregation (DLA). Deterministic fractals: (c) Vicsek and (d) Hexaflake. Deterministic lattices: (e) Kagome and (f) Hexagonal. In (a)-(f) we show: the number of particles, $N$; the radius, $\langle R\rangle$ or $R$; radius of gyration, $\langle R_g\rangle$ or $R_g$; fractal dimension, $D$ (estimated), and $D^*$ (theoretical); and the sixfold bond-orientational parameter, $\langle \Psi_6 \rangle$. The insets show the characteristic local structure.}
\label{fig:fig1}
\end{figure}

A visualization of the spatial systems here considered is presented in Fig. \ref{fig:fig1}. The physical structure of the systems is composed of identical particles forming connected structures via direct contact interactions. Each particle has a diameter equal to one unit leading to sets of non-overlapping disks with centers at one unit of distance or, equivalently, point distributions where the shortest distance between two neighboring points is one unit. For the network representations, undirected and unweighted networks are created under a geometric graphs approach: links are established pair-wise for all $N=1.5\times 10^4$ particles if the euclidean distance between two particles (nodes), $d_{ij}$, is equal or less than a given distance, $r$, this, is, $d_{ij}\leq r$, where $r$ is the euclidean connectivity parameter. For details of the systems' generation process and the mathematical definitions of the spatial quantities and network metrics used, please see Methods. 

\subsection*{Range, coverage, cost and configurational complexity ($r=1$)} 

\begin{figure}[t!]
\includegraphics[width=1.0\textwidth]{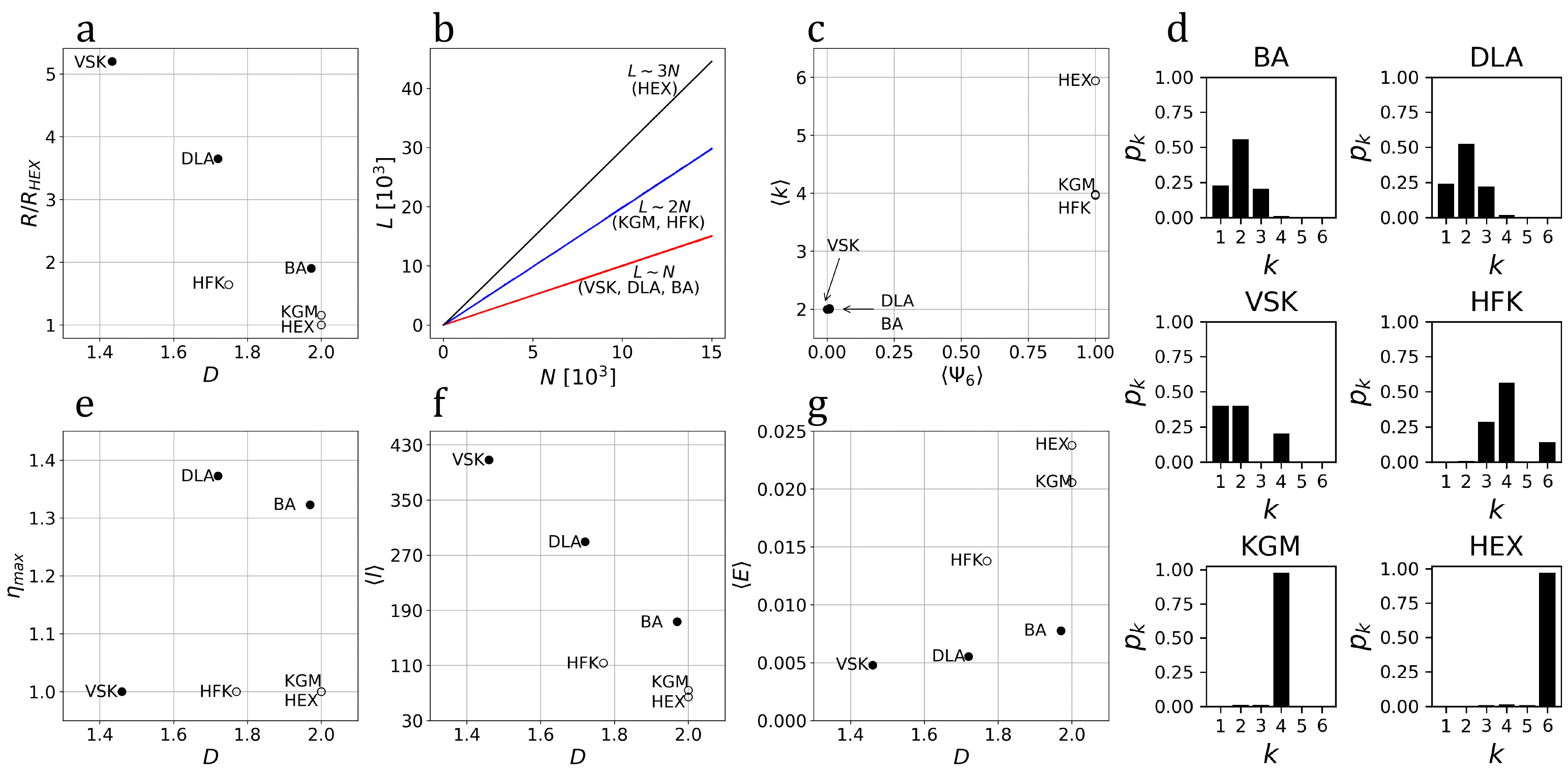}
\caption{\small \textbf{Structural Analysis ($r=1$).} (a) $R/R_{HEX}$ as a function of $D$. (b) The number of connections, $L$, as a function of $N$. (c) The average degree, $\langle k \rangle$, as a function of the sixfold bond-orientational parameter, $\langle \Psi_6 \rangle$. (d) Degree distribution, $p_k$, for each system. Network metrics as function of the fractal dimension, $D$: (e) the center-perimeter communication ratio, $\eta_{max}$, (f) average shortest-path length, $\langle l \rangle$, and (g) network efficiency, $\langle E \rangle$. }
\label{fig:fig2}
\end{figure}

Under the previous considerations, we first considered the case for $r=1$, which provides the exact network representation for direct contact interactions. The first important result is that for $r=1$, the physical branching morphology of BA and DLA corresponds to that of a tree network (no loops at the micro-level over all the structure \cite{vandewalle1997}). This result might seem self-evident or could be taken for granted due to a bias related to the cluster's physical macro shape, but in fact is not trivial. To formally prove this, we quantified the average clustering coefficient, $\langle C \rangle$, and found it to be technically zero for BA, DLA and even for the Vicsek fractal, as shown in Table \ref{tab:tab1}. One can observe in Table \ref{tab:tab1} that $\langle C \rangle$ is not exactly zero for BA and DLA, this is because for every $10^4$ aggregated particles it can be found a couple of them forming triangles with their neighbors due to the stochastic nature of the models. Furthermore, as it's well known, an important disadvantage of a tree network structure is that the network becomes disconnected very easily. Nevertheless, this fragility can be dealt with in many ways as it will be shown further on.

The characterization of the morphological features of the network and its capacity to explore the plane was done in terms of the \textsl{range}, \textsl{coverage}, \textsl{structural cost} and \textsl{configurational complexity}, given the same number of particles ($N=1.5\times 10^4$). The \textsl{range} is defined as the maximum euclidean distance with respect to the origin (linear extension), and the \textsl{coverage} is regarded as the ability to cover the plane in all directions (a space-filling property). The \textsl{range} is quantified by the characteristic radius, $R$, defined as the distance of the furthest particle in the cluster with respect to the origin averaged over the ensemble. For the \textsl{coverage}, we considered the fractal dimension, $D$, where $D\to 1$ would be associated to anisotropic linear structures, while $D\to 2$ to isotropic plane-filling structures. This dimension is calculated from the radius of gyration, $R_g$ (see Methods). The results for these quantities are indicated in Fig. \ref{fig:fig1} and Table \ref{tab:tab1}.

As expected, we found that the linearity induced by the branching morphology of tree-like systems (BA, DLA, and Vicsek) provides them with a range greater than that of the lattices (Hexagonal, Kagome, and Hexaflake), that, on the other hand, have the advantage of filling the plane more homogeneously due to their tight packing. Notably, the Hexagonal, Kagome and BA systems have the same dimension, $D=2$, despite having very different morphology. The Hexaflake and DLA fractals cover the plane with a similar scaling, $D=1.75$ and $D=1.73$, respectively, while Vicsek with $D=1.47$ reflects its more linear structure. These results are put together in Fig. \ref{fig:fig2}a, where we compare $R/R_{HEX}$ to the corresponding fractal dimension, $D$. Here, one can observe that, although lattice systems have a high coverage, they have a short range. On the other hand, tree-like structures are more versatile, they have a longer range and span different dimensions, including the one of the embedding space.

The \textsl{structural cost} is quantified in a first approximation by the number of contacts among particles or edges in the network, $L$. In Fig. \ref{fig:fig2}b, we show $L$ as function of $N$. All systems have a linear dependence, with $L\approx N$ for BA, DLA, and Vicsek; $L\approx 2N$ for Kagome and Hexaflake; and $L\approx 3N$ for the Hexagonal lattice. Notably, considering that $L=N-1$ for any tree-like system (such as BA, DLA, and Vicsek), then the number of connections required to build any tree-like system is independent of its fractal dimension.

As measures of \textsl{configurational complexity} we considered the degree distribution, $p_k$, average degree, $\langle k \rangle$, average sixfold bond-orientational parameter, $\langle \Psi_6 \rangle$, and average clustering coefficient, $\langle C \rangle$. We found that the average degrees of tree-like systems are statistically equivalent ($\langle k \rangle=2$), despite their very different global morphology. Similarly for the Kagome lattice and the Hexaflake fractal ($\langle k \rangle=4$) despite having very different fractal/non-fractal nature. Variations in local complexity were also quantified considering the average sixfold bond-orientational parameter, $\langle \Psi_6 \rangle$, which for ordered hexagonal systems, $\langle \Psi_6 \rangle =1$, whereas for non-hexagonal or disordered systems, $\langle \Psi_6 \rangle \to 0$ \cite{zangi1998, mazars2008, ledesma2021}. We found that for the Hexagonal, Kagome and Hexaflake systems $\langle \Psi_6 \rangle =1$, whereas for the BA, DLA, and Vicsek systems $\langle \Psi_6 \rangle = 0$, regardless of their fractal dimension. These results are put togther in Fig. \ref{fig:fig2}c. A more detailed description of local configurations is given by the degree distribution (see Fig. \ref{fig:fig2}d). For any particle in the plane, the degree, $k$, goes from one ($k_{min}=1$) to six ($k_{max}=6$). In this range, particles exhibit different local connectivity configurations, such as tips ($k=1$), branches ($k=2$) and bifurcations ($k=3$) for trees, and regular connections for the lattices. 

\subsection*{Center-perimeter communication and efficiency ($r \geq 1$)} 

\begin{figure}[ht!]
\centering
\includegraphics[width=1.0\textwidth]{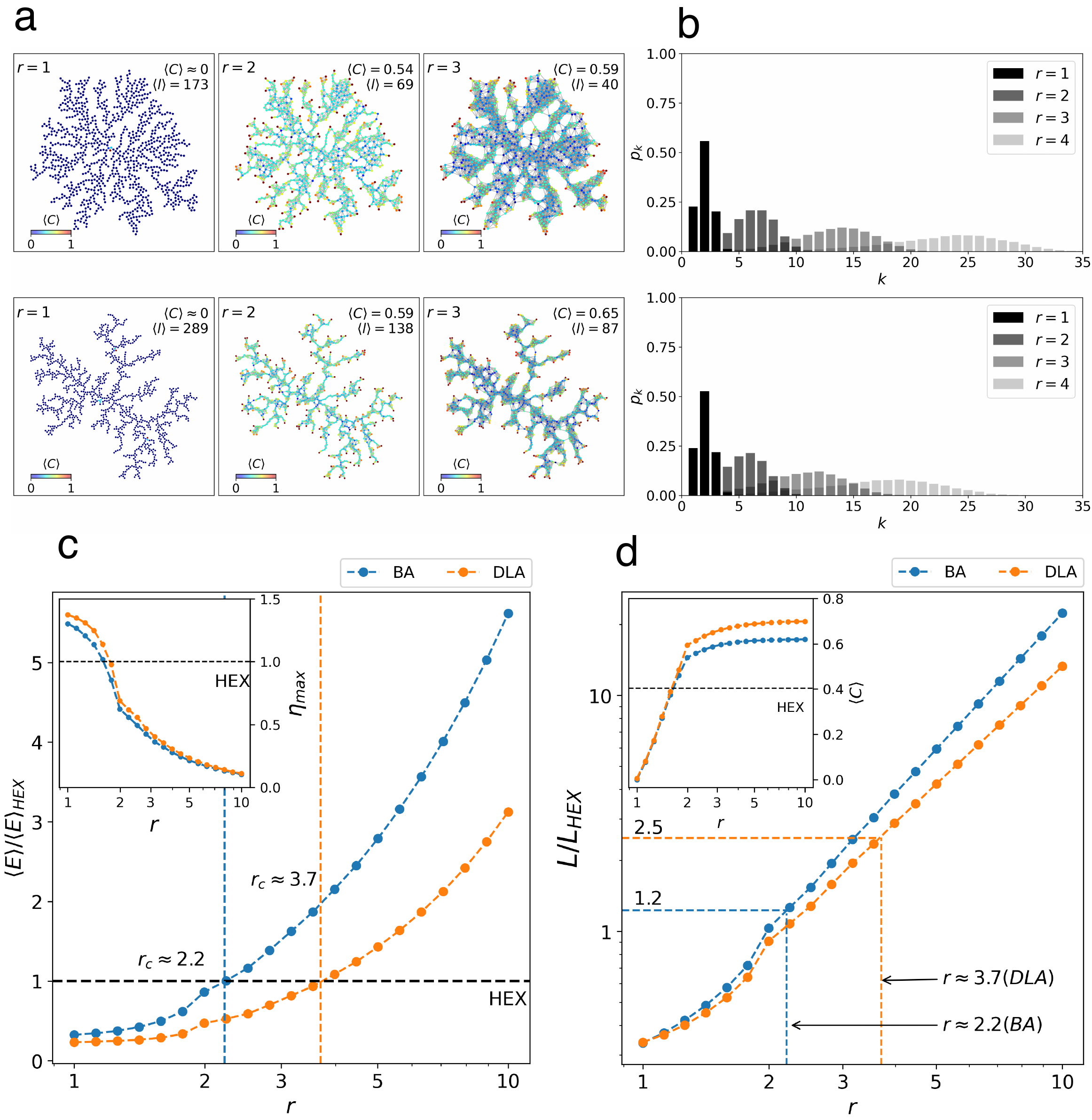}
\caption{\small \textbf{Efficiency analysis ($r\geq 1$)}. (a) Network visualizations of the BA and DLA networks for $r=\{1,2,3\}$ as indicated. Nodes are colored according to the clustering coefficient. (b) Corresponding degree distribution, $p_k$. Metrics as function of the connectivity parameter, $r$: (c) network efficiency, $\langle E \rangle/\langle E \rangle_{HEX}$, with the center-perimeter communication ratio, $\eta_{max}$, in the inset; (d) structural cost, $L/L_{HEX}$, with the clustering coefficient, $\langle C \rangle$, in the inset. In (c) and (d), the dotted horizontal line indicates the corresponding value of the metric for the Hexagonal lattice.}
\label{fig:fig3}
\end{figure}

We characterized the effectiveness of information exchange considering two perspectives: from center to perimeter and across the whole network. Considering first $r=1$, we quantified the network center-perimeter ratio, $\eta_{max}$, the average shortest-path length, $\langle l \rangle$, and the average communication efficiency, $\langle E \rangle$. The network center-perimeter communication ratio, $\eta_{max} = l_0^{max}/R$, where $l_0^{max}$ is defined as length of the path that connects the particle at the origin (center or source) of the system with the particle whose distance corresponds to $R$ (the perimeter). From Fig. \ref{fig:fig2}e, we found that $\eta_{max}>1$ for BA and DLA fractals, due to their stochastic branched morphology (center-perimeter paths highly deviate from a straight-line), whereas for deterministic fractals and the lattices $\eta_{max}=1$, given their ordered morphology (center-perimeter paths form straight-lines). For the average shortest-path length, $\langle l \rangle$ (see Fig. \ref{fig:fig2}f), we found that tree-like systems have long shortest-path lengths due to their branches and centralized structure, this is, in order for two particles in different branches to communicate, a path must go through the origin, whereas for lattice systems, the average shortest-path lengths are smaller due to their decentralized structure. This is consistent with the value of the average communication efficiency, $\langle E \rangle$, that depends on the inverse of all the pair-wise distances and considers that the farther two points in the network are, the less efficient they are to communicate \cite{latora2001}. Results for $\langle E \rangle$ as function of the fractal dimension, $D$, are shown in Fig. \ref{fig:fig2}g.

The efficiency of tree networks can be improved by adding connections among nodes beyond contact interactions or euclidean first-neighbors, which in turns affects the fragility of the tree structures, as previously explained. Following our geometric graphs approach, we introduced more links by allowing the value of the connectivity parameter to be greater than one unit, this is, $r\geq 1$. This criterion is uniformly applied pair-wise to all $N=1.5\times 10^4$ particles in the system. As it is show in Fig. \ref{fig:fig3}a, adding more links beyond first neighbors has a direct effect on quantities such as the clustering coefficient, which increases indicating more resilient or robust networks. Remarkably, although the BA and DLA fractals are known to be scale-free in space, they possess bounded degree distributions for different values of an euclidean connectivity parameter and, therefore, do not represent ordinary scale-free networks (see Fig. \ref{fig:fig3}b). As $r$ increases, the degree distributions broadens and distributes over a bounded degree range, however, as $r\to R$ the distribution will localize again, being exactly $p_k=1$ for $r=2R$.

Adding more links implies an additional structural cost. As such, we are interested in the cut-off value of the connectivity parameter, $r_c$, for which, the efficiencies of the BA and DLA fractals become equal or greater than the efficiency of the ordinary Hexagonal lattice at $r=1$, this is, $\langle E \rangle/\langle E \rangle_{HEX}=1$, and the structural cost that this entails. In Fig. \ref{fig:fig3}c we show the efficiency, $\langle E \rangle$, of the BA and DLA fractals relative to the efficiency of the Hexagonal lattice. As expected, the efficiency improves for $r>1$ due to the reduction of the all length paths in the network, including the one related to the center-perimeter measure (see inset in \ref{fig:fig3}c). Notably, the cut-off values occur at very small values of the connectivity parameter: BA at $r_c\approx 2$ and DLA at $r_c\approx 4$. At these cut-off values, the center-perimeter ratio, $\eta_{max}$, and the clustering coefficient, $\langle C \rangle$ (see inset in \ref{fig:fig3}d), are almost equal or even larger than the corresponding values of the Hexagonal lattice.

Regarding the structural cost, in Fig. \ref{fig:fig3}d we show the number of edges, $L$, relative to the number of edges of the Hexagonal lattice, $L_{HEX}$, as function of $r$. Considering the cut-off values, $r_c$, for BA and DLA, we found that the number of edges needed to build networks in which the BA or DLA are used as underlying spatial systems, with the same or better efficiency than the Hexagonal lattice, requires almost the same number of edges of an Hexagonal grid for BA ($L/L_{HEX}\approx 1$), and at least twice the number of edges for DLA ($L/L_{HEX}\approx 2.5$). This result can be understood by considering that $L$ increases exponentially as function of the connectivity parameter, $r$, thus, small changes in $r$ are sufficient to produce important changes in $L$.

\begin{table}[ht!]
\centering
\caption{\small \textbf{Numerical results of the spatial and network analysis for $r=1$.}}
\label{tab:tab1}
\begin{tabular}{|c|c|c|c|c|c|c|}
\hline
                               & \textbf{HEX} & \textbf{KGM} &  \textbf{HFK} & \textbf{BA}         &\textbf{DLA}           & \textbf{VSK} \\ \hline
\textbf{$N$}                   &   15000            &        15000    &      15000          &   15000             &   15000               &       15001     \\ \hline
\textbf{$R$}             &   64               &        74       &      106            & $122 \pm 1.27$      & $234 \pm 4.07$        &       334       \\ \hline
\textbf{$R_{g}$}               &   45               &        53       &      75             & $ 74 \pm 0.17 $     & $126 \pm 0.86 $       &       224       \\ \hline
\textbf{$D$}                   &   2.00             &        2.00     &      1.75           & $1.97\pm 0.015 $    & $1.72 \pm 0.0067$     &       1.47      \\ \hline
\textbf{$\langle\Psi_6\rangle$}&   1                &        1        &      1              &$0.0088 \pm 0.002$   & $0.0087 \pm 0.001$    &       0         \\ \hline
\textbf{$L$}                   &   44553            &        29744    &      29854          & $15006 \pm 0.89$    & $15092 \pm 2.59$      &       15000     \\ \hline
\textbf{$\langle k \rangle$}   &   5.9404           &       3.9658    &       3.9805	    & $2.0008\pm 0.0001$  & $2.0122 \pm 0.0003$   &       1.9998    \\ \hline
\textbf{$\langle C \rangle$}   &   0.4035           &       0.3382    &       0.4367        & $0.00049\pm0.00005$ & $0.00541 \pm 0.002$   &       0.0       \\ \hline
\textbf{$l_{0}^{max}$}         &   64               &       74        &      106            & $161.4 \pm 3.04 $   & $321.2 \pm 6.92$      &       334       \\ \hline
\textbf{$\langle l \rangle$}   &   64.2009          &       74.1370   &      113.3715       & $173.4341 \pm 0.38$ & $289.5194 \pm 1.19$   &       408.9502  \\ \hline
\textbf{$\langle E \rangle$}   &   0.0237           &       0.0205    &       0.0137        & $0.0077\pm 0.00001$ & $0.00553 \pm 0.00001$ &       0.00479   \\ \hline
\end{tabular} 
\end{table}

\section*{Discussion}

In the first part of the analysis, we characterized the systems' capacity to explore the plane in terms of the \textsl{range} (the maximum distance a system advances with respect to the origin), \textsl{coverage} (the ability to cover the plane in all directions), \textsl{structural cost} (assembly connections), and \textsl{configurational complexity} (local connectivity), given the same amount of resources (particles) under a geometric graphs approach for euclidean first-neighbors ($r=1$). We found that:

\begin{itemize}

    \item The local linearity induced by the branching morphology of tree-like systems (BA, DLA, and Vicsek) provides them with a range greater than that of the lattices (Hexagonal, Kagome, and Hexaflake).
    
    \item Tree structures are more versatile, they have a long range and span different dimensions, including those close to the dimension of the embedding space. Assuming that the amount of energy required to create any assembly connection is the same, the structural cost to create any tree structure is the same regardless of their fractal dimension. However, they are quite fragile (zero clustering).

    \item Small variations in the micro configurations lead to the emergence of very different macro structures; morphological macro-properties (tree-likeness and fractality) cannot be predicted from the micro-properties of the particles alone.
    
\end{itemize}

These results suggest that tree structures have the best balance between range, coverage and cost, they would be the best for the task of exploring space. However, the structural cost of building such networks would not only be associated to the amount of particles and connections (matter and energy), but also to the functionality under the uncertainty of the environment (lack of information). As such, information flow among any pair of points in the network, as well as from the center (or source) to the growing perimeter or any point the structure, is an important factor that weights-in in the cost. This is addressed in the second part of the analysis in terms of center-perimeter communication ratio and the network \textsl{efficiency} (overall network communication) under a geometric graphs approach beyond euclidean first-neighbors ($r\geq 1$). We found that:

\begin{itemize}
    
    \item Lattice systems have the best efficiency due to their decentralized structure (shorter length paths), whereas tree-like structures, despite their low structural cost, are inefficient due to their branched and centralized morphology (bigger length paths). This efficiency can be improved by adding links beyond contact neighbors, at a cost that can be similar to that of the Hexagonal lattice. This provides the same or even better improvements of the topological properties, such as the average shortest-path length, center-perimeter communication ratio and clustering (increase in robustness).

    \item Tree networks improved by adding a small (local) redundancy, have the best balance between range, coverage and cost, thus, they would be the best for exploring space without compromising the exchange of information over the entire network structure and by improving communication from center to the growing perimeter.

    \item Noteworthy, although the BA and DLA fractals are scale-free in space, they have bounded degree distributions for different values of the euclidean connectivity parameter and, therefore, do not represent ordinary scale-free networks.

\end{itemize}

Biological systems such as vascular networks \cite{ieva2014}, hyphal networks \cite{BOSWELL201230, dikec2020hyphal}, neurons \cite{ieva2014, smith2021neurons}, slime molds \cite{fessel2012}, and bacterial colonies \cite{benjacob1997}, display complex structures rich in branched or tree-like spatial features. The morphology of these systems seem to solve an adaptive exploration problem related to the maximization of the space that a connected structure can cover in order to retain or gain conditions for survival given limited amounts of matter, energy and information, and according to the demands of the environment \cite{fricker2009adaptive, bejan2013, bejan2017, torsten2020}. One characteristic of these systems is their physical fractality, often quantified by the fractal dimension \cite{ieva2014, vicsek2001fluctuations, lennon2015, nicolas2020}. However, although the fractal dimension is a good global measure of morphological complexity, it does not provide a comprehensive account of the micro structural features that could make branched morphologies relevant at the macro level for the system's biological function \cite{smith2021neurons, banavar1999, motoike2010, rocks2020}. The results obtained from our structural analysis based on fractal and network theory provide a detailed quantitative description and further insights into the morphological and topological features that make spatial systems with tree-like structure better at exploring space under limited amounts of resources and information (random growth in all directions under fixed number of particles), and the non-trivial interplay between the physical and topological properties of such complex systems. In addition, although much of the work has been devoted to the theory of random geometric graphs, our results contribute to the few research done on the properties of spatial networks with, or generated from, fractal spatial distributions.

An unambiguous characterization of complex spatial systems is still an open research problem. A good understanding of the fundamental aspects behind the development of such systems could provide valuable insights into medicine \cite{ieva2014, lennon2015, curtin2021}, engineering \cite{bejan2017, ziaei2015, demis2015}, biomimetic materials and biomorphs \cite{gruber2011biomimetics}, and sustainable cities \cite{fessel2012, kay2022stepwise, auerbach2021}. 

\section*{Methods}

\subsection*{Models}

Spatial networks are created from two-dimensional systems of identical particles forming connected structures with radial symmetry. In all the simulations, particles have a diameter equal to one unit leading to sets of non-overlapping disks with centers at one unit of distance or, equivalently, point distributions where the shortest distance between two neighboring points is one unit. All systems are centered at the origin and contain $1.5\times 10^4$ particles in order to have precise measurements of spatial quantities, such as the fractal dimension, as well as robust statistics for the network analysis (see Table \ref{tab:tab1}).

\begin{itemize}

    \item The \textsl{DLA fractal} emerges from the aggregation of particles moving under random trajectories upon contact with a static cluster \cite{witten1981}. We followed a standard procedure for particle-cluster aggregation in which particles are launched from a circle of radius $L = r_{max} + \delta$. Here, $r_{max}$ is the distance of the farthest particle in the cluster with respect to a seed particle at the origin and $\delta=100$ is used to avoid screening effects \cite{nicolas2016, nicolas2017}. In order to speed up the aggregation process, we also used a standard scheme that modifies the mean free path (set to one particle diameter) as the particles wander at a distances greater than $L$ or in-between branches, and set a killing radius at $2L$. The fractal dimension of DLA in the plane has been estimated at $D\approx 1.71$ \cite{nicolas2016, nicolas2017}.

    \item The \textsl{BA fractal} emerges from the aggregation of particles moving in ballistic or straight-line trajectories upon contact with a static cluster \cite{VOLD1963684}. We followed a standard procedure for particle-cluster aggregation in which particles are launched at random from the circumference of a circle of radius $L = r_{max} + \delta$. Here, $r_{max}$ is the distance of the farthest particle in the cluster with respect to a seed particle at the origin and $\delta=1000$ is used to avoid screening effects \cite{nicolas2016, nicolas2017}. The fractal dimension of BA in the plane has been estimated at $D\approx 2$ \cite{nicolas2016, nicolas2017}.

    \item The \textsl{Hexaflake fractal} is constructed iteratively by exchanging hexagons (scaled by a factor of 1/3) at the position of the vertices and centers of previous hexagons. These hexagons can also be replaced by points (one for each vertex, including the center) in such a way that the fractal can be generated using the following iterated function system \cite{mendivil2003fractals}:
        \begin{align*}
            w_0 &= \frac{1}{3}(x, y),\\
            w_k &= \frac{1}{3}(x + \cos\theta_k, y + \sin\theta_k) ,
        \end{align*}
    
    where $\theta_k=k\pi/3$ and $k=\{1,2,3,4,5,6\}$. As initial seed we considered $(x_0,y_0)=(1,1)$. There are $7^{n-1}$ points in the $n$-th iteration, each at a distance smaller by $1/3$ than in the previous iteration. In order to have a spatial distribution centered at the origin where the shortest distance between two neighboring points is one unit, all points are re-scaled using $(x^*,y^*)=3^n(x,y)-(1,1)$, where $(x,y)$ are the points at the $n$-th iteration. Its fractal dimension is given by $D=\log7/\log3\approx 1.771$.

    \item The \textsl{Vicsek fractal} is constructed iteratively by exchanging squares (scaled by a factor of 1/3) at the position of the vertices and centers of previous squares. These squares can also be replaced by points (one for each vertex, including the center) in such a way that the fractal can be generated using the same iterated function system of the Hexaflake, with $\theta_k=k\pi/2$ and $k=\{1,2,3,4\}$. As initial seed we considered $(x_0,y_0)=(1,1)$. There are $5^{n-1}$ points in the $n$-th iteration, each at a distance smaller by $1/3$ than in the previous iteration. In order to have a spatial distribution centered at the origin where the shortest distance between two neighboring points is one unit, all points are re-scaled using $(x^*,y^*)=3^n(x,y)-(1,1)$, where $(x,y)$ are the points at the $n$-th iteration. Its fractal dimension is given by $D=\log5/\log3\approx 1.465$.

    \item The \textsl{Kagome lattice} or trihexagonal tiling is composed of particles distributed on the plane forming equilateral triangles and hexagonal voids, with each particle in contact with 4 neighbors. For finite systems, particles at the border have less than 4 neighbors.

    \item The \textsl{Hexagonal lattice} is composed of particles distributed on the plane forming an hexagonal grid with each particle in contact with 6 neighbors. For finite systems, particles at the border have less than 6 neighbors.
        
\end{itemize}

In the following, due to the stochastic nature of the BA and DLA fractals, we considered an ensemble of 10 clusters which provides a good estimation of spatial quantities and very robust network metrics (see Table \ref{tab:tab1}). The data for each model is available online (see Data Availability). All computations were done using custom \textsl{Python} code and the \textsl{NetworkX} library \cite{networkx}.

\subsection*{Spatial Analysis}

\begin{itemize}

    \item The radius, $R$, is defined as the maximum (farthest) distance to the origin of the $N$ particles in the cluster, $R=\mathrm{max}\{r_i\}$, with $r_i^2=x_i^2+y_i^2$, where $(x_i,y_i)$ are the coordinates of the $i$th-particle.

    \item The radius of gyration, $R_g$, is defined as the root-mean-square distance of all the particles in the cluster \cite{vicsek1992, meakin1998},

     \begin{align} \label{eq:rg}
         R_g^2 = \frac{1}{N}\sum_{i=1}^N |\vec{r}_i-\vec{r}_{cm}|^2.
     \end{align}
 
    Clusters are centered at the origin and have radial symmetry, therefore, $\vec{r}_{cm}\to 0$. Also, for large enough $N$ ($>10^3$) \cite{nicolas2016}, the size scales with the number of particles as the power law, $R_g(N) \propto N^\beta$, and the clusters can be considered as self-similar fractals with a fractal dimension, $D =1/\beta$ \cite{vicsek1992, meakin1998}. This is computed from a linear fit in log-log scale.
    
    \item The average sixfold bond-orientational order parameter, $\langle \Psi_6 \rangle$, is given by \cite{zangi1998, mazars2008, ledesma2021},
    
    \begin{align}
        \langle \Psi_6 \rangle &= \frac{1}{N} \Biggl| \sum_{i=1}^{N} \frac{1}{k_i} \sum_{j=1}^{k_i} \exp(\mathrm{i}6\theta_{ij}) \Biggr|
    \end{align}
    
    where, $k_i$ is the number of first (contact) neighbors of the $i$-th particle, and $\theta_{ij}$ is the angle between the vector connecting the $i$-th particle with its $j$-th neighbor, with respect to an arbitrary axis (which in our analysis corresponds to the $x$-axis). For ordered hexagonal systems, $\langle \Psi_6 \rangle =1$, whereas for disordered systems, $\langle \Psi_6 \rangle \to 0$. 

\end{itemize}

\subsection*{Network Analysis}

We considered undirected unweighted networks \cite{menczer_fortunato_davis_2020, newman2018networks} created under a geometric graphs approach in which links are established pair-wise if the euclidean distance between two particles (nodes) is equal or less than the connectivity distance, $r$, this is, nodes $i$ and $j$ get connected if $d_{ij}\leq r$, where $d_{ij}=\sqrt{(x_i-x_j)^2+(y_i-y_j)^2}$.

\begin{itemize}

    \item The degree, $k_i$, of a node, $i$, is defined as the number of links or neighbors, and the average degree, $\langle k \rangle$, is the arithmetic mean, 
    
    \begin{align} \label{eq:avdeg}
        \langle k \rangle = \frac{1}{N}\sum_{i=1}^{N}k_i.
    \end{align}
    
    In addition, given the density, $d = L/L_{max}$, where $L_{max}=N(N-1)/2$ is the maximum number of possible links, and the total number of links, $L$, expressed as, $2L = \sum_{i=1}^{N}k_i$, the average degree can be rewritten as, $\langle k \rangle = 2L/N = d(N-1)$. Furthermore, for a network of $N$ nodes, the degree distribution, $p_k$, provides the probability of randomly finding a node with degree $k$, $p_k = N_k/N$, where $N_k$ is the number of degree-$k$ nodes. 
    
    \item The clustering coefficient, $C_i$, is a measure of the level to which the neighbors of node $i$ ($k_i > 1$) are neighbors among them as well (formation of triangles or triads). It is defined as, 
    
    \begin{align} \label{eq:clust}
        C_i = \frac{\tau_i}{\tau_{i,max}} = \frac{2\tau_i}{k_i(k_i - 1)}, 
    \end{align}
    
    where $\tau_i$ is the number of pairs of neighbors or triangles involving the node $i$. The maximum number of triangles of $i$, $\tau_{i,max}$, is the number of pairs formed by their neighbors $k_i$. The average clustering coefficient is the arithmetic mean,
    
    \begin{align} \label{eq:avclust}
        \langle C \rangle = \frac{1}{N}\sum_{i}C_i, 
    \end{align}
    
    where the nodes of degree $k<2$ are excluded of the mean.
    
    \item The distance between two nodes is defined as the minimum number of links that are traversed in a simple (not self-intersecting) path connecting two nodes. Such path is known as shortest path and its length as shortest-path length. As such, the average shortest-path length, $\langle l \rangle$, is the mean of the shortest-path lengths between pairs of nodes in the network, $l_{ij}$, defined as,
    
    \begin{align} \label{eq:avpath}
        \langle l \rangle = \frac{2}{N(N-1)}\sum_{i,j} l_{ij}. 
    \end{align}
    
    \item The efficiency measures the effectiveness of the exchange of information through the network. The efficiency, $E_{ij}$, between the nodes, $i$ and $j$, is defined as inversely proportional to the distance, $E_{ij} = 1/l_{ij}$. If there is no path connecting the nodes, $l_{ij}=\infty$, and $E_{ij} = 0$; but if they are connected, $l_{ij}\geq 1$, and the value of the efficiency is bounded between $0$ and $1$. The average efficiency is defined as \cite{latora2001}, 	

    \begin{align} \label{eq:aveff}
	    \langle E \rangle = \frac{1}{N(N-1)}\sum_{i\neq j}\frac{1}{l_{ij}}.
    \end{align}
    
\end{itemize}


\section*{Data Availability}

The spatial data used in this article is available in the following repository: \url{https://doi.org/10.6084/m9.figshare.12692009}


\bibliographystyle{ieeetr}
\bibliography{references}

\section*{Conflict of interests}

The authors declare that the research was conducted in the absence of any commercial or financial relationships that could be construed as a potential conflict of interest.

\section*{Authors contributions}

ACFO performed the computational analysis. JRNC supervised the computational analysis and defined the theoretical framework. JLCE supervised the overall research. All authors participated in the discussion, writing and approval of the final manuscript.

\section*{Acknowledgments}

Partial financial support by CONACyT Mexico through the grant A1-S-39909 is acknowledged. ACFO acknowledges a fellowship from CONACyT.


\end{document}